# Controlling Dual-Comb Soliton Motion inside a single Fiber Laser Cavity

Julia A. Lang[1], Sarah R. Hutter[2], Alfred Leitenstorfer[2], and Georg Herink[1]

[1] Experimental Physics VIII – Ultrafast Dynamics, University of Bayreuth, Bayreuth, Germany
[2] Department of Physics and Center for Applied Photonics, University of Konstanz, Konstanz, Germany

*Corresponding author: georg.herink@uni-bayreuth.de

**Abstract:** Ultrafast science builds on dynamic compositions of precisely timed light pulses[1] and evolving sequences of pulses are observed inside almost every mode-locked laser[2–6]. However, the underlying physics remains barely controlled and utilized until now. Here, we demonstrate the fast and deterministic control of soliton motion for the generation of programmable ultrashort pulse patterns from a dual-comb mode-locked Er:fiber laser. Specifically, we harness intra-cavity modulation of individual solitons and their laser-intrinsic dynamics to facilitate the rapid tuning of two interlaced soliton combs. Upon extra-cavity temporal recombination of both combs, we obtain reconfigurable pulse patterns at arbitrary delays. Using high-throughput real-time spectral interferometry, we resolve the short-range inter-soliton motion upon external stimuli and we demonstrate the high-speed sweeping of picosecond pump-probe-delays and programmable free-form trajectories. This work introduces a novel approach to soliton control and paves the way for ultrafast instruments at unprecedented high tuning, cycling and acquisition speeds.

Sequences of ultrashort laser pulses form the basis of ultrafast sciences[1]. Sweeping of temporal separations is critical for fast acquisition speeds and high detection sensitivity[2–4]. Previous real-time measurements of ultrafast laser dynamics uncovered that pairs of solitons, so-called "soliton molecules", rapidly evolve on temporal separations from nano- down to few-10 femtoseconds[5–9]. Applying external stimuli, i.e. pump power modulations, allows for controlling soliton motion to certain degrees, such as the switching between fixed soliton bound-states in Ti:sapphire, Erbium- and Thulium-doped-fiber lasers[10–14]. However, the separations of bound-states are dictated by laser-specific effects and can not or only slowly be adjusted[15,16,12]. While intracavity modulators and pulse shapers offer additional degrees of control[17,18], their slow temporal response typically affects entire groups of solitons simultaneously. Thus, soliton control is currently very limited in efficiency, speed and repeatability.

Alternatively, the fast scanning of optical delays without mechanically moving elements is opened-up by dual-comb laser spectroscopy or asynchronous optical sampling (ASOPS), combining the output of two laser oscillators with detuned repetition rates[19–21]. Dual-comb operation in a single cavity – based on directional, spatial, polarization or frequency multiplexing – eliminates the requirement of two laser sources and can significantly reduce overall complexity[22–24]. In addition, two detuned combs can be generated upon interaction with travelling acoustic waves[3,25]. However, temporal delays accumulate linearly over consecutive round-trips and always sweep over the full pulse repetition period. Thus, typical laser repetition periods in the nanosecond range dictate long time delay windows and result in inefficient acquisition for ultrashort intervals in the pico- and femtosecond range. Electronically controlled optical sampling (ECOPS) offers tunable delay windows[26,27], but at the expense of two separate laser sources.

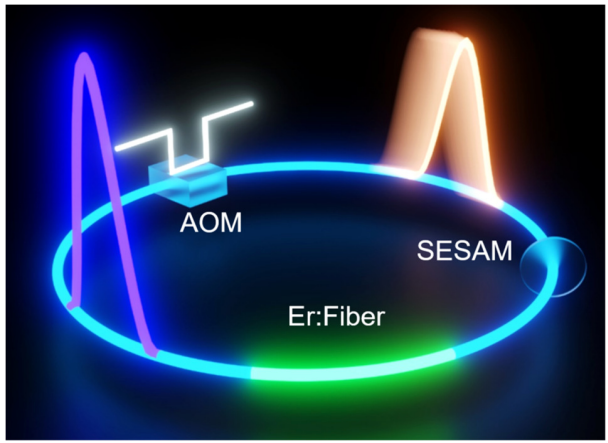

*Figure 1:* Concept of controlling dual-comb soliton motion: Intra-cavity acousto-optic modulation inside a Er:fiber laser acts on individual solitons from one comb and generates tailored inter-soliton trajectories.

In this work, we demonstrate the generation of ultrashort pulse sequences with rapidly swept and freely-programmable temporal delays by harnessing laser-intrinsic soliton dynamics. We implement comb-selective soliton control via intracavity modulations of individual pulses within two interlaced frequency combs in a single all-fiber laser, as illustrated in Figure 1. In the experiment, we generate two interlaced soliton combs in a second-harmonic mode-locked SESAM-based Er:fiber laser[28] with fundamental repetition rate of $f_{rep} = \frac{1}{T_{rep}} = 27$ MHz. Figure 2a displays the dual-comb source based on a unidirectional ring cavity. We introduce the intra-cavity soliton-control via a fiber-coupled acousto-optic modulator (AOM) allowing for fast intensity modulations of the zero-order beam down to a single pulse in-between both combs. The driving radio-frequency (RF) signal is synchronized to the harmonically mode-locked combs and allows for AOM bandwidths up to 200 MHz with modulation windows down to 10 ns. At the output, both interlaced combs can be combined via an adjustable asymmetric fiber-optic Mach-Zehnder interferometer at arbitrary delays $\Delta\tau = 0..T_{rep}/2$. We resolve the resultant relative soliton motion down to single roundtrips (RT) with a real-time oscilloscope via both direct photodetection and spectral interferometry for sub-nanosecond separations.

**Results**

First, we present experimental soliton trajectories for macroscopic delays in the nanosecond range. We apply modulations to one comb (orange coloured pulse in Fig. 2) repetitively, and we monitor the evolution of the pulse trajectories via real-time photodetection. Each frame captures four consecutive pulses and is triggered by the unmodulated comb (purple coloured pulse in Fig. 2). Starting the modulation, the solitons rapidly change separations and reach a new stable equilibrium separation, as displayed in Figure 2b. After switching-off the external stimulus, the system relaxes back to the harmonically mode-locked state with two equally spaced combs.

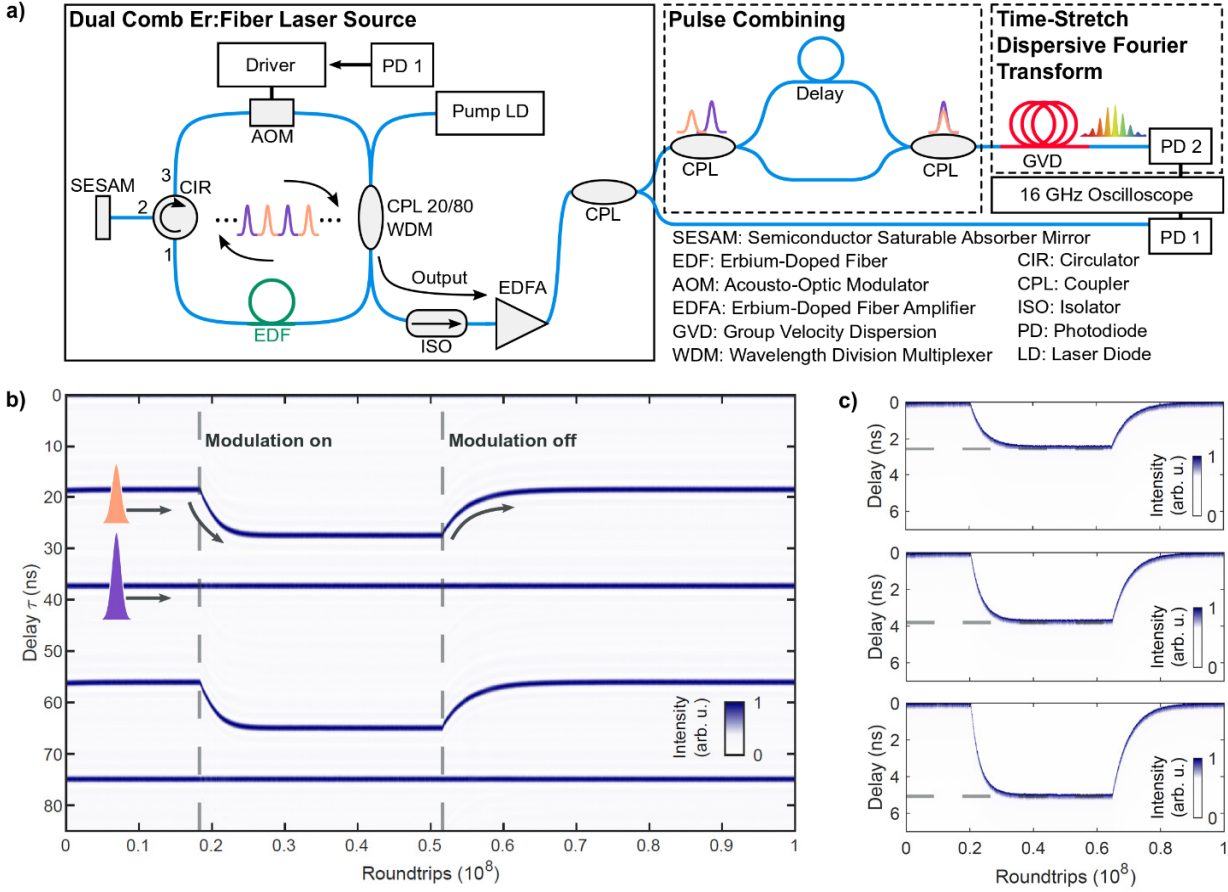

*Figure 2: a) Setup of the dual-comb fiber laser oscillator, optional external pulse combination and real-time detection. b) Experimental real-time trajectory of dual-comb motion upon transient modulation of one interlaced soliton comb via the AOM. c) Three trajectories for increasing modulation (by duty cycle) demonstrate the increase in relative soliton velocity and the tuneability of modulated equilibrium separations.*

The impact of modulation strength i.e. controlled via the duty cycle (higher number of modulations of one comb per time) is presented by the trajectories in Figure 2c): Increasing the effective modulation strength accelerates the motion and shifts the equilibrium separations. From the start of the motion, we find that the relative velocity follows a linear dependence with 1.4 fs/RT per 1% intensity difference. The intensity difference results from an interplay of accumulated AOM-induced loss and dynamic laser gain, as analysed and discussed further below. These sequences generate highly deterministic inter-soliton motion and enable rapid tailoring of soliton delays.

We now introduce a physical model underlying the soliton interactions and corroborate our experimental observations with numerical simulations. Our approach exploits the coupling of soliton intensity to group velocity. In this laser source, the coupling is provided predominantly by the SESAM. The transient saturable absorption process reshapes the pulse envelope asymmetrically in time and effectively delays the pulse via increased absorption at the pulse front[29] (sketched in Fig. 3a). The temporal shift is intensity-dependent due to the underlying saturation nonlinearity. Thus, intensity differences between both solitons are translated into relative temporal shifts. This behaviour is illustrated by two pulses of different intensities in Figure 3b. Based on numerical simulations for typical SESAM parameters, we find relative shifts on the order of ~1 fs/RT for a 1% intensity difference (see Supplementary Information).

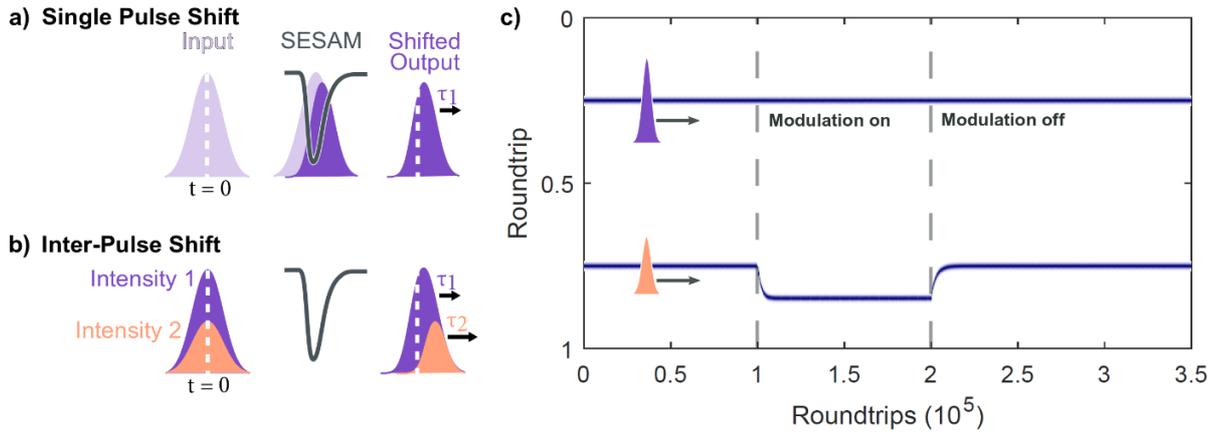

*Figure 3: a) Temporal shift due to the reshaping of pulses via saturable absorption. Initial centre of the pulse indicated by white dashed line. b) The nonlinear absorption induces different temporal shifts for different pulse intensities, resulting in intensity-dependent relative motion. c) Simulation of the relative pulse shifts between two harmonically mode-locked soliton combs upon intensity modulation in the presence of saturable absorption and transient laser gain: The modulated second pulse shifts and stabilizes at a new equilibrium separation. Switching the modulation off, the pulse returns to the initial harmonically mode-locked state.*

Moreover, the overall evolution of the soliton motion is governed by transient intensity differences due to laser gain dynamics. The saturated laser gain introduces a long-range coupling between multiple solitons since the gain does not fully recover within one roundtrip. This gain depletion and recovery effect provides the mechanism for harmonic mode-locking[30,31]: At the equidistant separation both combs experience identical gain, reach identical intensity and propagate at identical group velocity. The pulse-selective intensity modulation breaks this balance: The reduced intensity delays the pulse via the SESAM. For increasing delays, however, the gain recovers and fully compensates the continuous modulator loss at the new equilibrium separation. A simplified model evaluates the pulse shifts due to SESAM- and gain-induced reshaping in the presence of laser gain dynamics and reproduces the observed relative soliton trajectories (see Supplementary Information). We include the intensity modulation of one comb over repeated roundtrips and display the soliton trajectories in Figure 3c). In correspondence to the experimental acquisition, the fast time axis of individual frames is referenced to the unmodulated soliton. Upon activation of the AOM, the second comb further delays and approaches the new stable equilibrium. Switching-off the modulation, the harmonically mode-locked state is re-established.

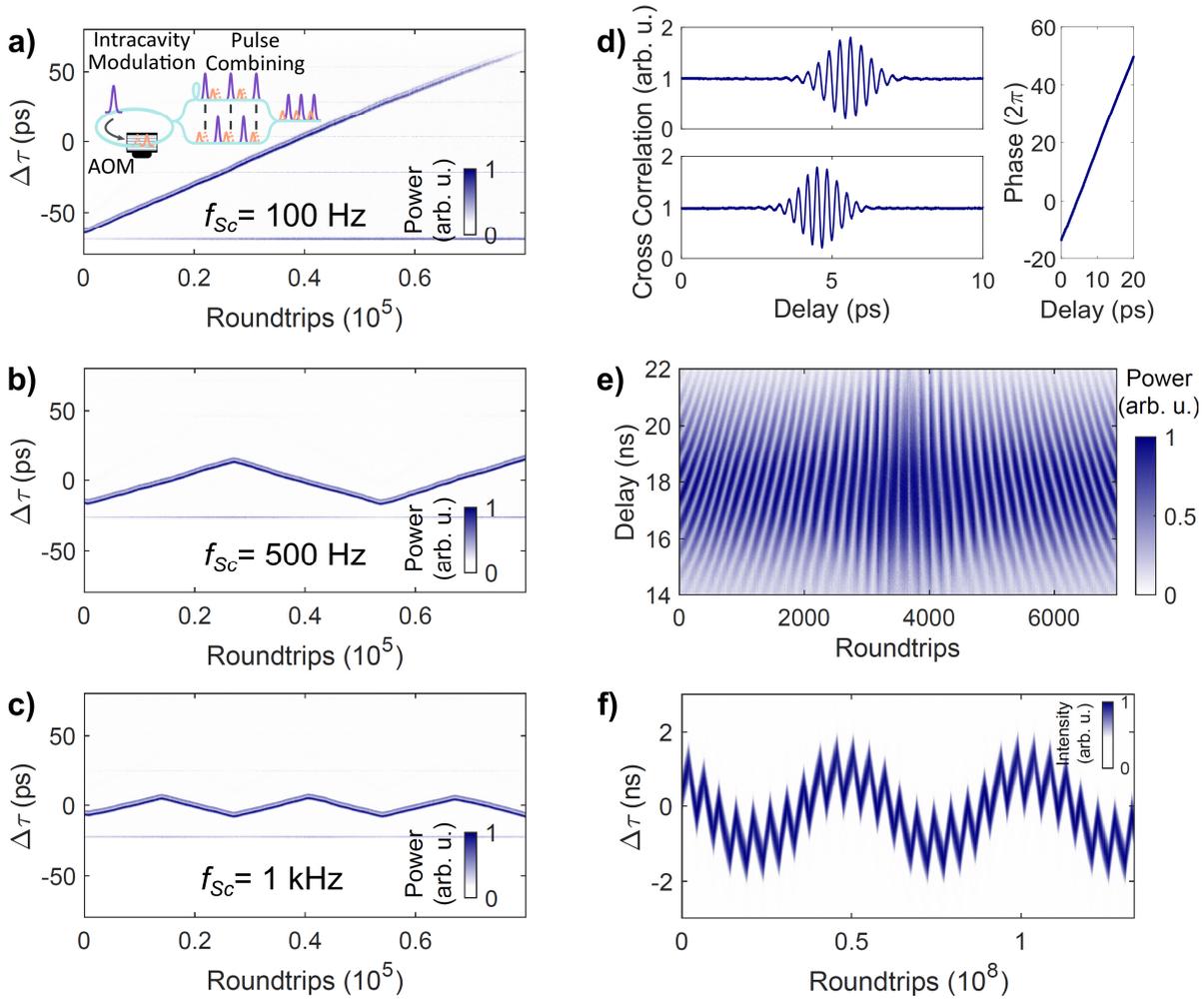

*Figure 4: a-c)* Scanning picosecond soliton separations. Both combs are temporally combined in an interferometer (inset). For increasing scanning rate, shorter delay windows are generated. *d)* Cross-correlations ($f_{Sc}$ = 800 Hz, downscans) evidence the interference between both combs upon temporal overlap. Picosecond-shifts between two successive interferograms depend on the time between two scans, i.e. scanning frequency. The interferometric oscillation period arises from the distinct relation between relative phase and delay (right), obtained from real-time spectral interferometry, shown for the upper trace in *e)*. *f)* A more complex free-form trajectory, i.e. a double-modulated sweep in the nanosecond-range, is obtained by variation of the scanning frequency.

Next, we focus on ultrafast soliton motion on sub-100 ps timescales below the temporal resolution of the oscilloscope. Experimentally, we employ real-time spectral interferometry based on time-stretch dispersive Fourier transformation[32,33] which allows for recording single-shot optical spectra for every laser roundtrip. In order to generate closely-spaced pulse pairs and to obtain spectral interference between both soliton combs, we split the laser output using a fiber-based Mach-Zehnder interferometer, introduce a variable delay in one arm and overlap two successive pulses upon recombination at the output, as sketched in Fig. 2a) and 4a). For example, delaying one arm with $T_{rep}/2$ generates the exact temporal overlap of both combs. We control the short-range soliton motion by switching repetitive modulations (at fixed duty cycle) of one comb on and off with a scanning frequency $f_{Sc}$, and we observe a regular linear scanning of delays, as displayed in Figure

4a)-c). The delays are measured via the Fourier transformation of real-time spectral interferograms[7], yielding linear cross-correlations. Since the relative velocity difference is fixed by the duty cycle of the modulation, a higher scanning frequency $f_{Sc}$ reduces the scanning delay window. For example, we achieve a scanning range of 12.5 ps at a frequency $f_{Sc}$ of 1 kHz.

By recording the laser output with a fast photodetector, we can directly measure cross-correlations and coherence properties of both combs. Figure 4d) displays two consecutive interferograms for downscans (modulation off) and reveals high interference contrast between both combs. The background of the second, non-overlapping pulse pair is subtracted and the traces are normalized to this corrected background. Strikingly, the interferometric oscillation period of 0.35 ps is ~67 times larger than the optical cycle. We directly obtain the relation between changes in relative phase and separation from spectral interferograms[7] (inset in Fig. 4d)). Depending on modulation, the carrier-envelope-frequencies of both combs change and, effectively, reduce the number of interference fringes. The shift between the zero-delay position of both interferograms results from ps-fluctuations accumulated between the two scans (see Supplementary Information for characterization of timing jitter). Such fluctuations depend on the laser design and operational parameters, and further active and passive means for stabilization can be applied. Finally, we also illustrate the possibility to generate programmable free-form soliton trajectories: Figure 4f) displays real-time measurements obtained via temporally-varying the scanning frequency, yielding double-modulated soliton trajectories.

**Discussion**
In summary, we introduce the control of dual-comb soliton motion inside a single fiber laser cavity for the programmable generation of soliton pulse patterns. We demonstrate the switching and continuous sweeping of pulse pairs from picosecond to nanosecond delay ranges. Currently, the approach allows for all-optical scanning of multi-picosecond delays at frequencies above 1 kHz and, thus, is suitable for rapid optical pump-probe and Raman spectroscopies[34]. In perspective, the selective intra-cavity modulation of single solitons offers novel strategies to probe multi-soliton interactions, is applicable to a wide range of fiber and solid-state laser systems, and may open-up a novel class of real-time instrumentation[35].

**Materials and Methods**
We implement the Er:fiber laser as a ring cavity with an optical circulator arm to ensure unidirectional propagation and to incorporate the SESAM mode-locking element. By adjusting the pump power, the laser is operated in the harmonically mode-locked state at a second harmonic repetition rate of 54 MHz. The spectral widths of 10 nm in the fundamental operation mode supports 260-fs pulses. Intracavity soliton-control is facilitated by incorporating a fiber-coupled acousto-optic modulator in zero-order transmission with a temporal modulation window down to 10 ns. The driving radio-frequency signal is synchronized to the harmonically mode-locked combs and allows for modulation up to 200 MHz. At the output, both interlaced combs can be combined via an adjustable asymmetric fiber-optic Mach-Zehnder interferometer at arbitrary delays $\Delta \tau = 0..T_{rep}/2$. We resolve the resultant relative soliton motion with a real-time oscilloscope (Tektronix DPO71604SX) with 16 GHz bandwidth, fast InGaAs photodetectors with bandwidths >15 GHz, and optional time-stretch dispersive Fourier transformation via a single-mode fiber with dispersion D = 990 ps/nm. Depending on the required resolution, we acquire successive frames separated by waiting periods between 250 ns and 520 µs. For the cross-correlation measurement, we employ an InGaAs photodiode with a bandwidth of 1 MHz.

**Funding:** Deutsche Forschungsgemeinschaft (project number 461131168).

**Author contributions:** J.A.L. performed the experiment and simulations, S.R.H. and A.F. fabricated the laser oscillator, J.A.L. and G.H. conceived the study, analysed the data wrote the manuscript. All authors contributed to the discussions.

**Data Availability:** Data underlying the results presented in this paper are available from the authors upon reasonable request.

**Competing interests:** All other authors declare they have no competing interests.

**Supplementary Materials**